\documentclass[pra,twocolumn,superscriptaddress,10pt]{revtex4-1}
\usepackage{graphicx,epsfig,epstopdf}
\usepackage[dvipsnames]{xcolor}
\usepackage{amsmath}
\usepackage{subfigure}
\usepackage{mathrsfs}
\usepackage{bm}
\usepackage{times}

\begin{document}

\title{Noncyclic nonadiabatic holonomic quantum gates via shortcuts to adiabaticity}

\author{Sai Li}

\author{Pu Shen}

\author{Tao Chen}
\affiliation{Guangdong Provincial Key Laboratory of Quantum Engineering and Quantum Materials,
and School of Physics\\ and Telecommunication Engineering,
	South China Normal University, Guangzhou 510006, China}

\author{Zheng-Yuan Xue}\email{zyxue83@163.com}
\affiliation{Guangdong Provincial Key Laboratory of Quantum Engineering and Quantum Materials,
and School of Physics\\ and Telecommunication Engineering,
South China Normal University, Guangzhou 510006, China}

\affiliation{Guangdong-Hong Kong Joint Laboratory of Quantum Matter,
and Frontier Research Institute for Physics,\\
South China Normal University, Guangzhou 510006, China}

\date{\today}

\begin{abstract}
  High-fidelity quantum gates are essential for large-scale quantum computation. However, any quantum manipulation will inevitably affected by noises, systematic errors and decoherence effects, which lead to infidelity of a target quantum task. Therefore, implementing high-fidelity, robust and fast quantum gates is highly desired. Here, we propose a fast and robust  scheme to construct high-fidelity holonomic quantum gates for universal quantum computation based on resonant interaction of three-level quantum systems via shortcuts to adiabaticity. In our proposal, the target Hamiltonian to induce noncyclic non-Abelian geometric phases  can be inversely engineered  with less evolution time and demanding experimentally, leading to high-fidelity quantum gates in a simple setup. Besides, our scheme is readily realizable in physical system currently pursued for implementation of quantum computation. Therefore, our proposal  represents a promising way towards fault-tolerant geometric quantum computation.
\end{abstract}

\maketitle

\section{Introduction}
Quantum computation, based on the laws of quantum mechanics,  is promising in solving hard computational problems and has many potential applications in modern science and technology \cite{Feynman}. However,  noises, systematic errors and decoherence effects will inevitably affect a quantum gate operation, which can result in the deviation of the target evolution, i.e., infidelity of a quantum gate. Thus, for  large scale quantum computation, noise-resistant and high-fidelity quantum gates are highly desired. Remarkably, geometric phases \cite{berry, fw,aa} only depend on the global properties of their evolution trajectories, and thus  can naturally be used to induce quantum gates that are robust against certain local noises  \cite{AN1,AN2,AN3,AN4,AN5,AN6}.

Recently, nonadiabatic holonomic quantum computation (NHQC), based on the non-Abelian geometric phase, has been proposed for high-fidelity quantum operation \cite{NJP, TongDM}, and then  significant theoretical \cite{surface1, Singleloopxu, Singleloop, SingleloopSQ, surface2, ChenAn, eric, surface3} and experimental progresses \cite{Abdumalikov2013,Feng2013, Zu2014,Arroyo-Camejo2014,nv2017, nv20172, li2017,Xu18, ni2018, kn2018, dje2019} have been made. However, the above NHQC implementations are sensitive to the systematic error \cite{Zheng16, Jing17} caused by external control imperfections, which can lead to infidelity of the implemented gate. The main reason for this sensitivity is due to  the fact that, in three-level quantum systems, the two coupling strengths in the above implementations need to be synchronized. This strong requirement leads to the fact that the two couplings are actually one independent variable, similar to the case of Rabi oscillation.

In implementing of NHQC, to improve the gate robustness against systematic control errors, various protocols have been proposed with preliminary experimental demonstrations. As the first stage, the conventional encoding methods have been proposed  \cite{TongDM,encode, zhangj2014, liang2014,zhouj, xue1, wangym2016, xue2, xue3, zhaopz2017, xue4, wangym2020}, which require more resources of physical qubits.  Then, other quantum control techniques are  introduced in cooperating with NHQC, such as the composite scheme or dynamical decoupling strategy \cite{composite, zhu2019, dd}, the deliberately optimal pulse control technique \cite{liubj17, Liu18, yan2019, Li, ai2020, ai2021, GPC, DCGLi}, and complex pulses target to shorten the gate-time \cite{xugf2018, zhang2019, Chentoc3, BNHQC, yuyang}, etc. However, these kinds of enhancement of gate robustness either need deliberate control of experimental pulse or greatly lengthen the gate-time. Therefore, improving the holonomic gate robustness in NHQC  with simplified control is highly desired.

Here, to further improve gate robustness in NHQC, we propose a fast and robust  scheme to construct high-fidelity holonomic quantum gates based on resonant interaction of three-level quantum systems via shortcuts to adiabaticity  (STA) \cite{STA2010b,STA2019}. Specifically, in our scheme, non-synchronous coupling strengths of the Hamiltonian for the target  noncyclic holonomic quantum gates can be inversely engineered with the help of the Lewis-Riesenfeld invariant (LRI) \cite{LR,Chen12,xia2018,zhou19,Yan19,IR2020}. Comparing with previous NHQC schemes, our scheme has the following merits. First, we only employ the conventional pulse shaping technique, which is experimentally friendly and can be readily   realized in current experiments. Second, arbitrary optimized gate can be achieved with smaller pulse area, which can lead to shorter gate time and thus smaller decoherence-induced gate infidelity. Besides, the gate robustness is greatly enhanced via numerically evaluation. In addition, our scheme can be readily implemented in conventional physical systems for quantum computation, e.g., superconducting quantum circuits, nitrogen-vacancy centres and trapped ions, etc. Therefore, our proposal  represents a promising way towards fault-tolerant quantum computation.

\section{Hamiltonian engineering via STA}

We now introduce how to construct the Hamiltonian for realizing a target evolution  based on the LRI theory \cite{LR,Chen12,xia2018,zhou19,Yan19,IR2020}. Here, we consider  a  $\Lambda$-type three-level system that is resonantly driven by two external fields, inducing the  coupling between auxiliary state $|e\rangle$ and qubit states $|0\rangle$, $|1\rangle$, respectively, as shown in Fig. 1(a). In addition, our protocol  can also be applicable in the V- or $\Xi$-type configurations. Assuming $\hbar=1$ hereafter, under the rotating wave approximation, the driven system can be expressed in the subspace $\{|0\rangle,|e\rangle,|1\rangle\}$ as
\begin{equation}\label{Hamiltion0}
\mathcal{H}(t)=\frac{1}{2}\left[\begin{array}{ccc}{0} & {\Omega_{0}(t)} & {0} \\ \Omega_{0}(t)& {0} & {\Omega_{1}(t) e^{-i \phi}} \\ {0} & {\Omega_{1}(t) e^{i \phi}} & {0}\end{array}\right],
\end{equation}
where $\Omega_k(t)$ with $k=0,1$ represents time-dependent coupling strength between auxiliary state $|e\rangle$ and qubit state $|k\rangle$, and {${\phi}$ is a constant relative phase}.
In general, this time-dependent Hamiltonian $\mathcal{H}(t)$ is difficult to be solved, and thus it is hard to obtain the evolution operator by directly using $U = \hat{{\mathcal{T}}}e^{-i\int^T_0\mathcal{H}(t)dt}$. Thus, the LRI theory  \cite{LR,Chen12,xia2018,zhou19,Yan19,IR2020} has been proposed to find exact solution of the time-dependent Schr\"{o}dinger equation of $\mathcal{H}(t)$, and thus the corresponding evolution operator  can be easily found. In the case of Hamiltonian $\mathcal{H}(t)$, there exists a LRI \cite{Yan19}
\begin{equation}\label{invariant}
I(t)=\frac{ \lambda}{2}\left[\begin{array}{ccc}{0} & {\cos \gamma \sin \beta} & {-i \sin \gamma e^{-i \phi}} \\ {\cos \gamma \sin \beta} & {0} & {\cos \gamma \cos \beta e^{-i {\phi}}} \\ {i \sin \gamma e^{i \phi}} & {\cos \gamma \cos \beta e^{i \phi}} & {0}\end{array}\right],
\end{equation}
where $\lambda$ is a constant, and $\gamma$ and $\beta$ are time-dependent variables designed {for a target task}. Furthermore, by solving $dI/dt \equiv\partial I/\partial t+(1/i)[I(t), \mathcal{H}(t)]=0$, constraint conditions between the coupling strength $\Omega_k(t)$ and the invariant parameters can be obtained as
\begin{eqnarray}\label{relationship}
\Omega_{0}(t)=2[\dot{\beta} \cot \gamma(t) \sin \beta(t)+\dot{\gamma} \cos \beta(t)],\notag \\
\Omega_{1}(t)=2[\dot{\beta} \cot \gamma(t) \cos \beta(t)-\dot{\gamma} \sin \beta(t)],
\end{eqnarray}
where $\dot{\beta}$ and $\dot{\gamma}$ denote the time derivative of ${\beta(t)}$ and ${\gamma(t)}$, respectively.

\begin{figure}[tbp]
	\begin{center}
		\includegraphics[width=8cm]{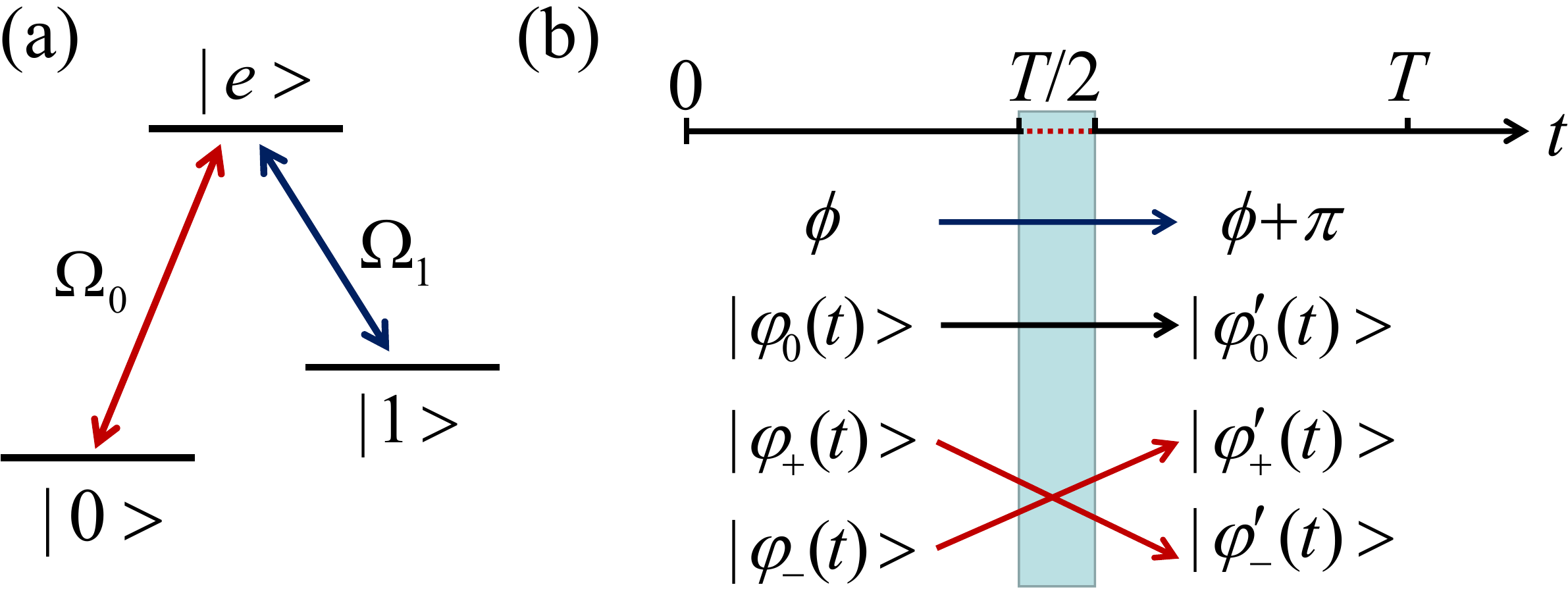}\label{fig1}
\caption{ (a) The coupling configuration for a three-level system to induce holonomic quantum gates. (b) The schematic show of the construction of the evolution path, where the  change of the constant phase factor from $\phi$ to $\phi + \pi$ of the Hamiltonian in Eq. (\ref{Hamiltion0}) at  $T/2$ leads to the exchange between the two orthogonal channels $|\varphi_\pm (t)\rangle$, which is the key to keep the final state staying within the qubit basis.}
	\end{center}
\end{figure}

In addition, eigenstates of $I(t)$ can be chosen as,
\begin{equation}\label{eigenstates0}
\left|\varphi_{0}(t)\right\rangle=\left[ \begin{array}{c}{\cos \gamma(t) \cos \beta(t)} \\ {-i \sin \gamma(t)} \\ {-\cos \gamma(t) \sin \beta(t) e^{i \phi}}\end{array}\right],
\end{equation}
and
\begin{equation}\label{eigenstates1}
\left|\varphi_{ \pm}(t)\right\rangle=\frac{1}{\sqrt{2}} \left[ \begin{array}{c}{\sin \gamma(t) \cos \beta(t) \pm i \sin \beta(t)} \\ {i \cos \gamma(t)} \\ {[-\sin \gamma(t) \sin \beta(t) \pm i \cos \beta(t)] e^{i \phi}}\end{array}\right],\\
\end{equation}
as a set of orthogonal dressed states, with the corresponding eigenvalues being $0$ and $\pm\lambda/2$, respectively. As an arbitrary evolution state $|\psi(t)\rangle$ under the driven Hamiltonian $\mathcal{H}(t)$ can be expanded as a linear superposition of dressed states as $|\psi(t)\rangle=\sum_{n=0, \pm} C_{n} e^{i \alpha_{n}}\left|\varphi_{n}(t)\right\rangle$ with $C_n$ being time-independent amplitudes, according to the LRI theory, the corresponding {LR phase} is {$\alpha_{n}(t)=\int_{0}^{t}\left\langle\varphi_{n}\left(t^{\prime}\right)\left|[i ({\partial}/{\partial t^{\prime}})-\mathcal{H}\left(t^{\prime}\right)\right]|\varphi_{n}\left(t^{\prime}\right)\right\rangle d t^{\prime}$}, which is the global phase accumulated on a dressed state. This LR phase includes both geometric (first term) and dynamical (second term) parts. Then, the time-evolution operator is
\begin{equation}\label{Operator}
U(t)=\sum_{n=0, \pm}e^{i \alpha_{n}}\left|\varphi_{n}(t)\right\rangle\langle \varphi_{n}(0)|,
\end{equation}
where $\alpha_{\pm}(t) = \mp\int_{0}^{t}{\dot{\beta}(t^{\prime})}/{\sin\gamma(t^{\prime})}  dt^{\prime}$ and $\alpha_{0}=0$.
That is, according to the required properties of $\alpha_n$, after properly designing {time-dependent variables ${\beta}$, ${\gamma}$ and the time-independent ${\phi}$}, a target Hamiltonian $\mathcal{H}(t)$ can be inversely engineered according to Eq. (\ref{relationship}), and the resulting time-evolution operator of the qubit system is determined by Eq. (\ref{Operator}).

\section{Holonomic quantum gates}

In this section, we present the construction of fast and robust {{holonomic}} quantum gates based on the non-belian geometric phase, {by applying the strategy of inverse-engineering Hamiltonian described above.} Meanwhile, the gate performance is numerically evaluated, which proves that our scheme is more robust than {the conventional NHQC scheme} \cite{Singleloop}, under  same conditions.

\subsection{Gate construction}
We first proceed  to the holonomic  gate construction.A
For the purpose of removing dynamical phase and realizing pure non-Abelian geometric evolution, we deliberately divide the evolution path into two equal parts, changing the driving parameters at the intermediate moment as shown in Fig. 1(b).
To be specific, in the first part $t \in [0, \textrm{T}/2]$, the boundary conditions are set to be $\phi_1=\phi$, $\gamma(0)=\gamma(\textrm{T}/2)=0$, $\beta(0)=\theta/4$ and $\beta(\textrm{T}/2)=0$. By these settings, according to Eq. (\ref{Operator}), the evolution operator of the first part is
\begin{equation}\label{Operator1}
U(\textrm{T}/2,0)=\sum_{n=0, \pm}e^{i \alpha_{n}(\textrm{T}/2)}|\varphi_{n}(\textrm{T}/2)\rangle\langle \varphi_{n}(0)|.
\end{equation}
Then, in the second part $t \in [\textrm{T}/2, \textrm{T}]$, {to make sure that the target states still stay in the qubit subspace and the evolution time is as {short} as possible, we  exchange the two evolution channels $|\varphi_{+}(t)\rangle$ and $|\varphi_{-}(t)\rangle$ by changing $\phi$ to $\phi + \pi$ in the intermediate moment. Besides, we also target to set $\alpha_{\pm}(\textrm{T})= - \alpha_{\mp}(\textrm{T}/2)$ to fully {eliminate} the dynamical phase.} Thus, the boundary conditions can be chosen as $\phi_2=\phi+\pi$, $\gamma(\textrm{T}/2)= \gamma(\textrm{T}) = 0$, $\beta(\textrm{T}/2) = 0$ and $\beta(\textrm{T}) =  \theta/4$, resulting in the evolution operator for the second part as
\begin{equation}\label{Operator2}
U(\textrm{T},\textrm{T}/2)=\sum_{n=0, \pm}e^{i \alpha_{n}(\textrm{T})}|\varphi^\prime_{n}(\textrm{T})\rangle\langle \varphi^\prime_{n}(\textrm{T}/2)|.
\end{equation}
In this way, during the  whole {evolution time {$\textrm{T}$}}, the non-Abelian geometric evolution operator within the quit subspace $\{|0\rangle, |1\rangle\}$ is
\begin{equation}\label{final}
\begin{aligned}
U(\textrm{T})&= U(\textrm{T},\textrm{T}/2) U(\textrm{T}/2,0)\\
&= \left( \begin{array}{cc} \cos\frac{\theta} {2}  & -\sin\frac{\theta} {2}e^{-i \phi}\\
 \sin\frac{\theta} {2}e^{i \phi} &   \cos\frac{\theta} {2}
\end{array}\right),
\end{aligned}
\end{equation}
which can be regarded as rotation around axis $\cos\phi\sigma_y-\sin\phi\sigma_x$ by an angle $\theta \in [0,\pi]$ denoted as $R[\theta,\phi]$. Thus, by choosing different $(\theta,\phi)$, universal holonomic single-qubit gates can be realized.

\begin{figure}[tbp]
	\begin{center}
		\includegraphics[width=8cm]{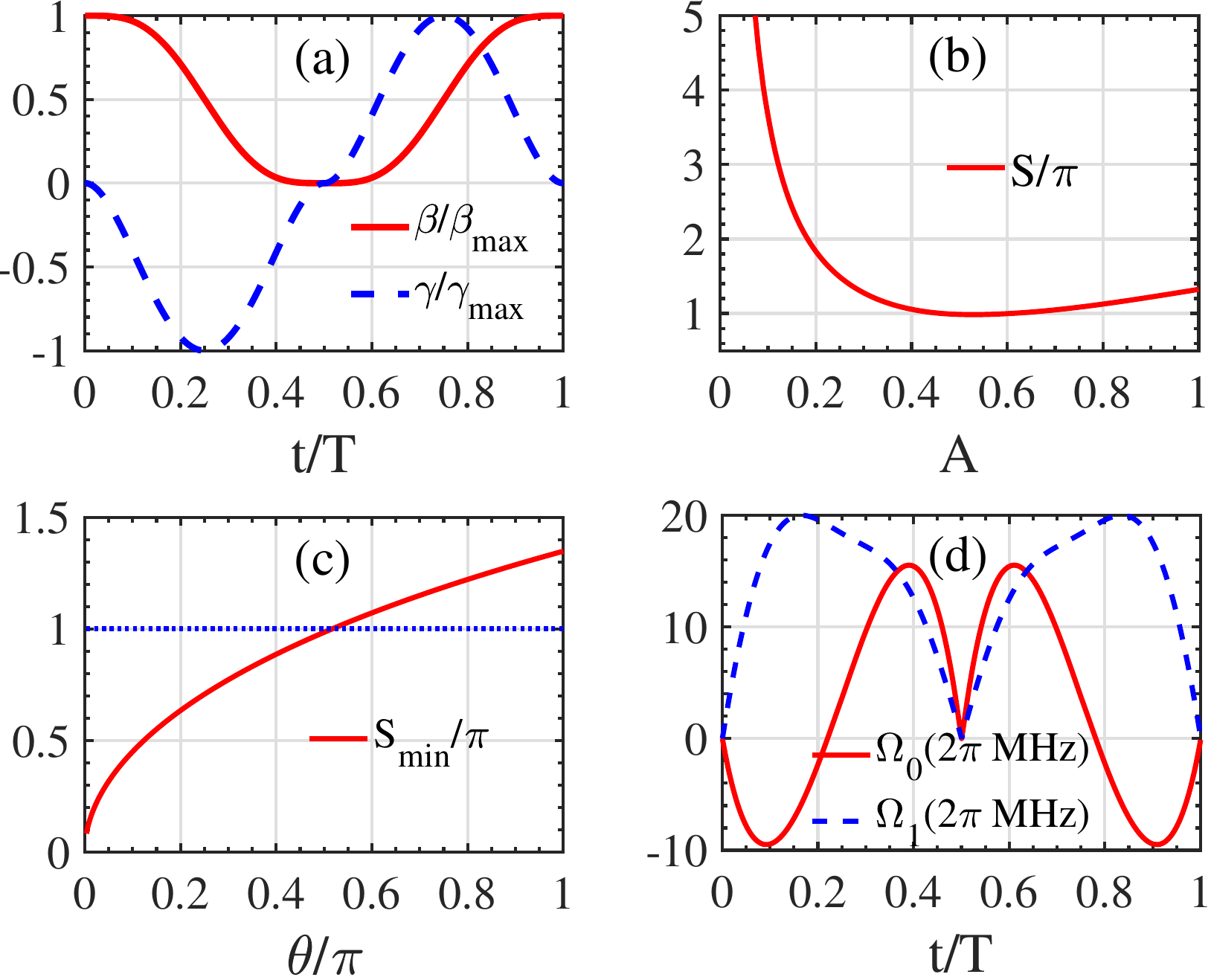}
		\caption{(a) The shapes of the two time-dependent  variables of $\gamma$ and $\beta$. (b) The pulse area $S$ with respect to $A$ for $\theta = \pi/2$. {(c) The minimum pulse area $S_{min}$ with respect to $\theta$ (red solid line) and the blue dashed line indicates the case of conventional NHQC. (d) The slopes of the coupling strength $\Omega_k$ for the Hamiltonian in Eq. (\ref{Hamiltion0})  when $\theta = \pi/2, A = 0.46, \textrm{T}=29.5$ ns and $S=\pi$.}}\label{fig2}
	\end{center}
\end{figure}

\subsection{Gate performance}

Here, we choose {a proper set of time-dependent variables}, which  satisfy the boundary conditions \cite{zhou19}, to show the performance of the implemented holonomic quantum gates. That is, in the first time interval $t\in [0, \textrm{T}/2]$, we set
\begin{eqnarray}\label{GB1}
\gamma(t) &=& -\frac{A}{(\textrm{T}/4)^4} t^2(t-\textrm{T}/2)^2 , \notag\\
\beta(t) &=& 35\theta\left[\frac{(\textrm{T}/2-t)^4 }{4(\textrm{T}/2)^4}
 - \frac{3(\textrm{T}/2-t)^5 }{5(\textrm{T}/2)^5}\right.\notag\\
 &&+ \left. \frac{(\textrm{T}/2-t)^6 }{2(\textrm{T}/2)^6}
 -\frac{(\textrm{T}/2-t)^7}{7(\textrm{T}/2)^7}   \right],
\end{eqnarray}
where {$A$ is a tunable parameter}. And, in the second time interval $t \in [\textrm{T}/2, \textrm{T}]$, we  set
\begin{eqnarray}\label{GB2}
\gamma(t) &=& \frac{A}{(\textrm{T}/4)^4} (t-\textrm{T}/2)^2(t-\textrm{T})^2 ,\notag\\
\beta(t) &=& 35\theta\left[\frac{(t-\textrm{T}/2)^4 }{4(\textrm{T}/2)^4} -\frac{3(t-\textrm{T}/2)^5 }{5(\textrm{T}/2)^5} \right. \notag\\
&&\left. + \frac{(t-\textrm{T}/2)^6 }{2(\textrm{T}/2)^6}
- \frac{(t-\textrm{T}/2)^7}{7(\textrm{T}/2)^7} \right].
\end{eqnarray}
The above choice of the boundary conditions can ensure the time-dependent variables to be simple and realizable, and the shapes of parameters $\gamma(t)$ and $\beta(t)$ are shown in Fig. 2(a).
Then, the corresponding form of $\Omega_k$ in Hamiltonian $\mathcal{H}(t)$ can be reversely engineered  according to Eq. (\ref{relationship}). {Notably, we can find that the pulse area $S$, defined by $S = \frac{1}{2}\int^\textrm{T}_0\sqrt{\Omega^2_0+\Omega^2_1}dt$, of a fixed rotation operation is related to the tunable parameter $A$ directly. Therefore, for a certain rotation angle $\theta$, the minimum pulse area $S_{\textrm{min}}$ can be obtained by  adjusting $A$, as show in Fig. 2(b) for the gate with $\theta = \pi/2$.} Moreover, as shown in Fig. 2(c), we also plot  the minimum pulse area $S_{\textrm{min}}$ with respect to rotation angle $\theta$.  In addition, we can obtain the average gate-time for all possible rotation angle, which indicate that the average gate-time of our scheme is shorter compared with the conventional NHQC case \cite{Singleloop} (i.e., $S\equiv\pi$), resulting  in higher gate fidelity.

\begin{figure}[tbp]
	\begin{center}
		\includegraphics[width=8cm]{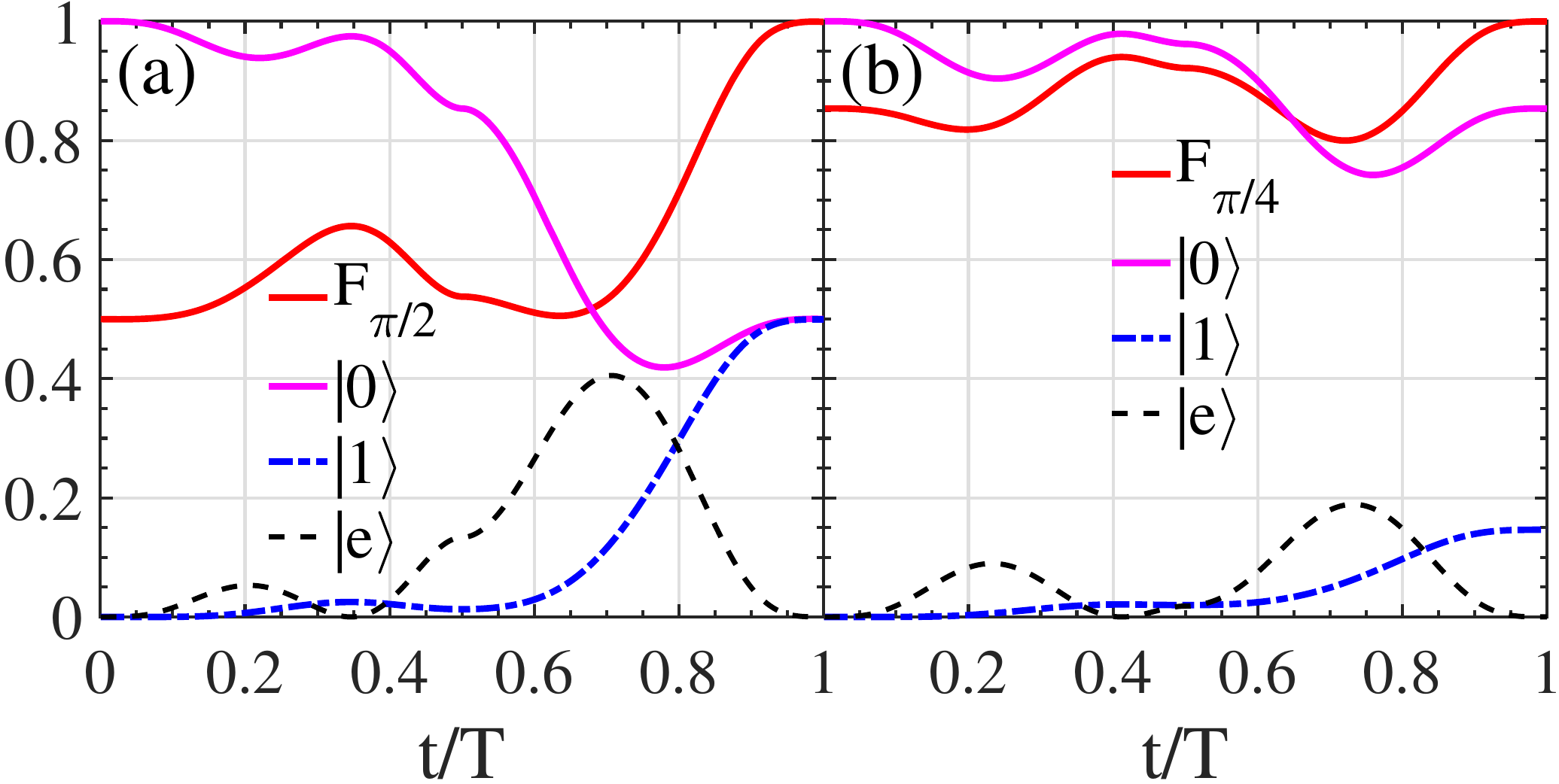}
\caption{The dynamics of the state population and fidelity for  (a) $R[\pi/2,\pi/2]$  and (b)  $R[\pi/4,\pi/2]$ gate operations. }\label{fig3}
	\end{center}
\end{figure}

To  evaluate the performance of the constructed holonomic gates, we use the Lindblad master equation of
\begin{eqnarray}
\label{master}
\dot\rho &=& i[\rho, \mathcal{H}(t)]  + \frac{1}{2}  \left[ \Gamma_1 \mathcal{L}(\sigma_1) + \Gamma_2 \mathcal{L}(\sigma_2)\right],
\end{eqnarray}
where $\rho$ is the density matrix of the qubit system and $\mathcal{L}(\mathcal{A})=2\mathcal{A}\rho \mathcal{A}^\dagger-\mathcal{A}^\dagger \mathcal{A} \rho -\rho \mathcal{A}^\dagger \mathcal{A}$ is the Lindbladian of operator $\mathcal{A}$; {$\sigma_1=|0\rangle\langle e|+|1\rangle\langle e|$ and $\sigma_2=(|e\rangle\langle e|-|0\rangle\langle 0|)+(|e\rangle\langle e|-|1\rangle\langle 1|)$} with $\Gamma_1$ and $\Gamma_2$ being the corresponding decay and dephasing rates, respectively. Here, considering the energy cost \cite{Cost1, Cost2} and experimental feasibility, we set the maximum coupling strength to be $(\Omega_k)_{\textrm{max}}=2\pi \times 20$ MHz. In addition, the decay and dephasing rates are set as $\Gamma_1=\Gamma_2= 2\pi \times 5$ kHz. {For two typical gate operations $R[\pi/2,\pi/2]$ and $R[\pi/4,\pi/2]$, the minimum pulse area $S_{\textrm{min}}$  can be obtained with $A = 0.46$ and $0.38$, respectively.} Assuming that  the quantum system is initially prepared in $|\psi_1\rangle=|0\rangle$, the obtained gate-fidelities of $R[\pi/2,\pi/2]$ and $R[\pi/4,\pi/2]$ are as high as $F _{\pi/2}= 99.91\%$ and $F_{\pi/4}= 99.97\%$, as shown in  Figs. 3(a) and 3(b), respectively, where the state fidelity is defined by $F_{\theta}=\langle\psi_{f_\theta}|\rho|\psi_{f_\theta}\rangle$ with $|\psi_{f_\theta}\rangle=R[\theta,\pi/2]|0\rangle$ being ideal final state.
In addition, for a general initial state $|\psi_1\rangle=\cos\theta^{'}|0\rangle+\sin\theta^{'}|1\rangle$, gate operations $R[\pi/2,\pi/2]$ and $R[\pi/4,\pi/2]$ will result in ideal final state $|\psi_{f_\theta}\rangle=\cos\theta^{'}(\cos(\theta/2)|0\rangle +i\sin(\theta/2)|1\rangle)+\sin\theta^{'}(-i\sin(\theta/2)|0\rangle + \cos(\theta/2)|1\rangle)$. To fully evaluate the performance of these two gates, the gate-fidelity can be defined as $F^G_{\theta}=\frac{1}{2\pi}{\int^{2\pi}_0}\langle\psi_{f_\theta}|\rho_1|\psi_{f_\theta}\rangle d\theta^{'}$ \cite{Poyatos97}, where the integration is numerically performed for 1001 input states with $\theta^{'}$ being uniformly distributed over $[0,2\pi]$, and we obtain the fidelities of the two gates as $F^G _{\pi/2}= 99.84\%$ and $F^G _{\pi/4}= 99.92\%$, respectively.

\begin{figure}[tbp]
	\begin{center}
		\includegraphics[width=8cm]{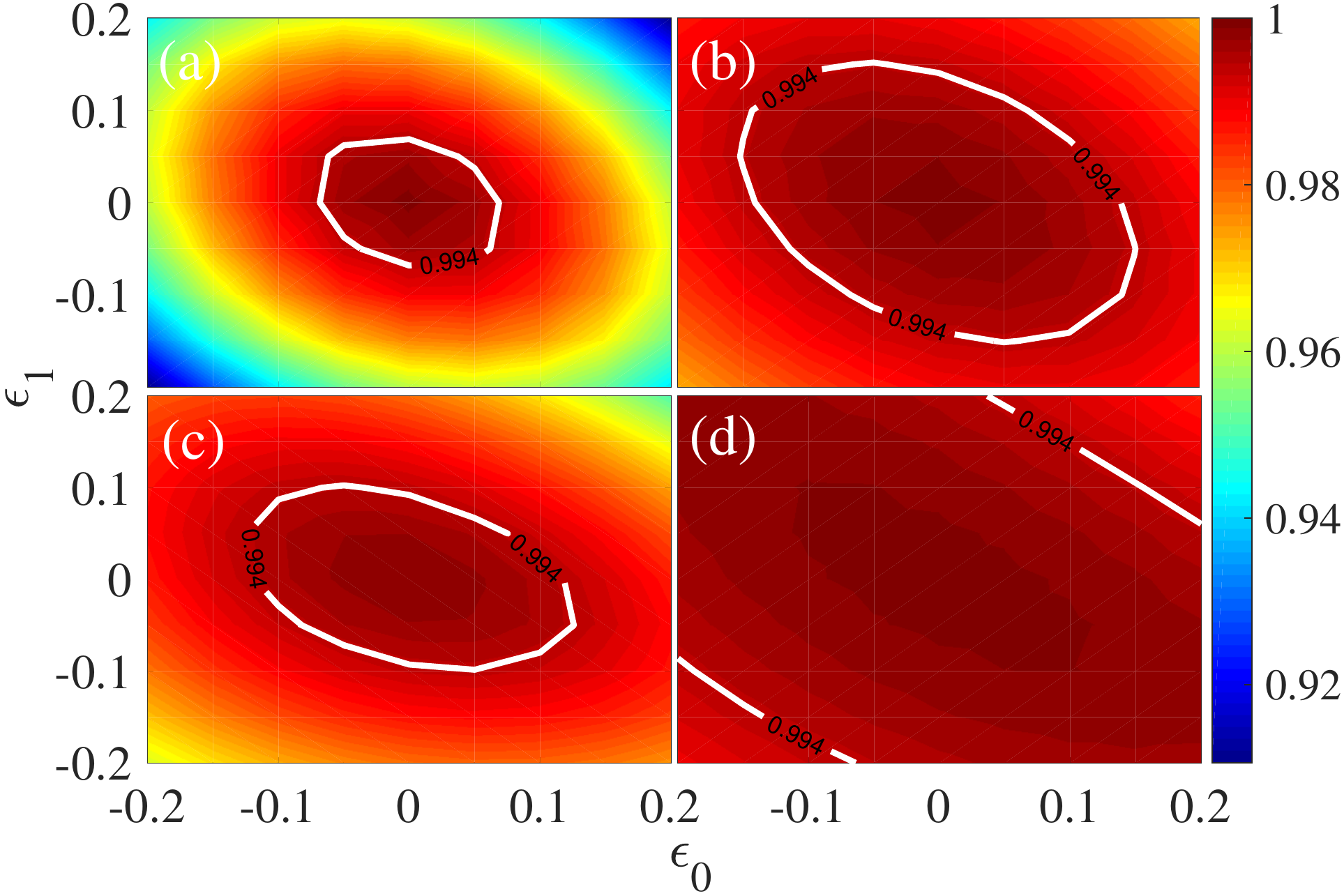}\label{fig4}
		\caption{Gate fidelities under systematic Rabi errors. (a) and (b) describe the fidelities of the $R_X[\pi/2]$ and $R_X[\pi/4]$ gates in the NHQC case. (c) and (d) present  the fidelities of $R[\pi/2,\pi/2]$ and $R[\pi/4,\pi/2]$ gates in our noncyclic scheme. }
	\end{center}
\end{figure}

\subsection{Gate robustness}
Different to the conventional NHQC case, the holonomic gates here are implemented with non-synchronous coupling strengths $\Omega_k$. {Therefore, we further numerically simulate the gate robustness of these two cases, against the static systematic Rabi errors.} Under this error, the interaction Hamiltonian  $H(t)$ in Eq. (\ref{Hamiltion0}) changes  to
\begin {equation}\label{Hamiltion1}
\mathcal{H}^\prime(t)=\frac{1}{2}\left[\begin{array}{ccc}{0} & (1+\epsilon_0){\Omega_{0}} & {0} \\ (1+\epsilon_0)\Omega_{0}& {0} & (1+\epsilon_1){\Omega_{1} e^{-i \phi}} \\ {0} & (1+\epsilon_1){\Omega_{1} e^{i \phi}} & {0}\end{array}\right],
 \end {equation}
where $\epsilon_k$ are the static systematic Rabi error fractions. In the following, we use Hamiltonian $\mathcal{H}^\prime(t)$ and the Lindblad master equation in Eq. (\ref{master}) to evaluate the robustness of the quantum gates for our scheme and the conventional NHQC scheme.

\begin{figure}[tbp]
	\begin{center}
\includegraphics[width=8cm]{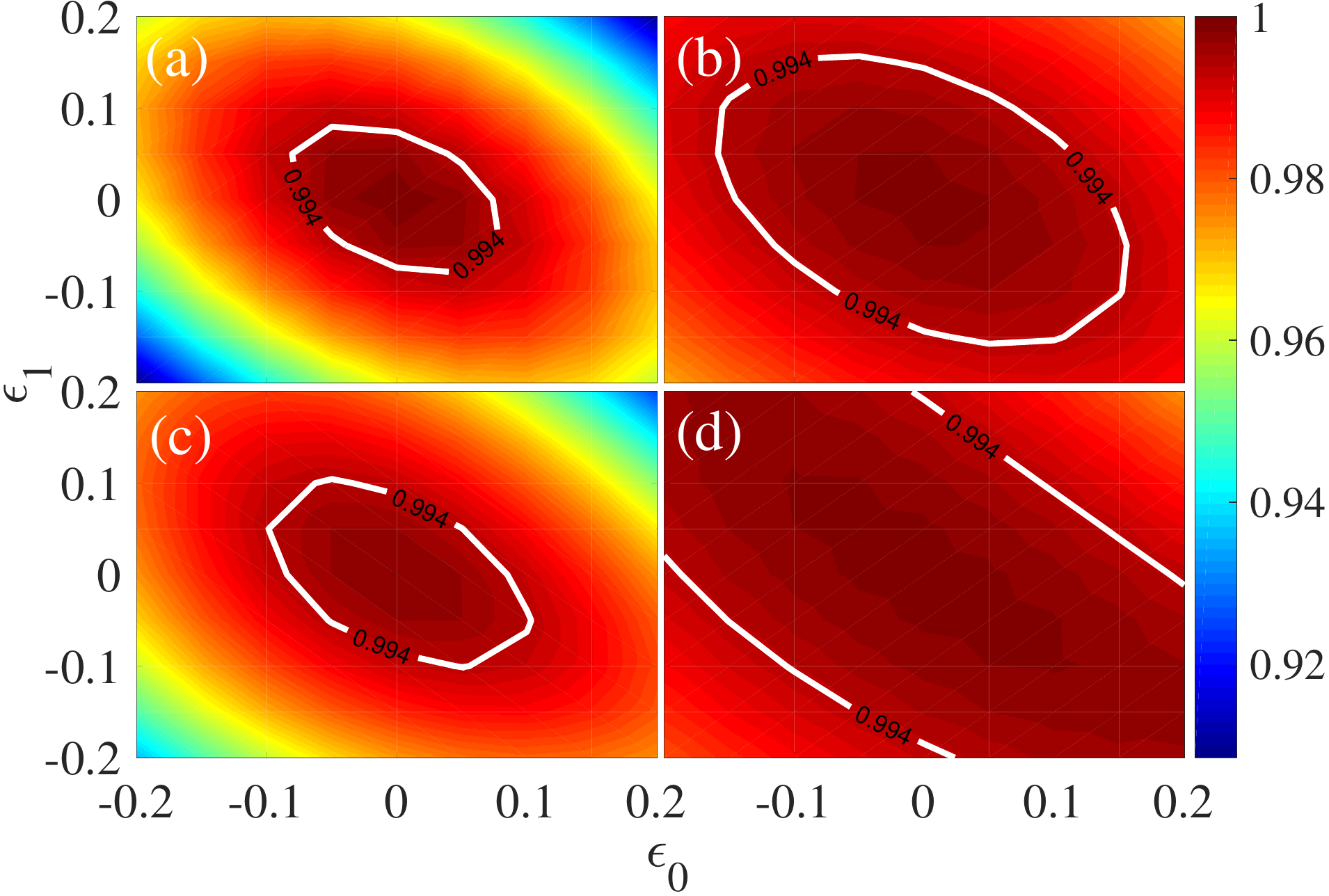}\label{fig5}
\caption{Gate fidelities under systematic Rabi errors.  (a) and (b) describe the fidelities of the $R_Y[\pi/2]$ and $R_Y[\pi/4]$  gates in the
NHQC case. (c) and (d) present the fidelities of $R[\pi/2,0]$ and $R[\pi/4,0]$  gates in our noncyclic scheme.}
	\end{center}
\end{figure}

{To compare our proposal with the NHQC case under the same conditions,} we simulate gate performance in the NHQC case with the same pulse shape of $\Omega_1$, as shown in Fig. 2(d). Furthermore, the evolution operator of our scheme $R[\theta,\pi/2]$ is coincident  with $R_X[\theta]$ of the NHQC case, which is a rotating operator around the $\sigma_x$ axis with an angle $\theta$.  As shown in Fig. 4(a) and Fig. 4(b),  we simulate the fidelities of  $R_X[\pi/2]$ and $R_X[\pi/4]$ gates for the NHQC case, with $-0.2\leq\epsilon_k\leq0.2$, respectively. And the fidelities of  $R[\pi/2,\pi/2]$ and $R[\pi/4,\pi/2]$ gates in our scheme are shown in Fig. 4(c) and Fig. 4(d). It clearly indicates that our scheme is more robust than the NHQC case. Besides, $R[\theta,0]$ corresponds to $R_Y[\theta]$, which is the rotation around the $\sigma_y$ axis. And the numerical simulation results   of  the  $R_Y[\pi/2]$ and $R_Y[\pi/4]$ gate-fidelities are shown in Fig. 5(a) and Fig. 5(b), respectively. For comparison purpose, as shown in Fig. 5(c) and Fig. 5(d), gate-fidelities of  the evolution operator $R[\pi/2,0]$ and $R[\pi/4,0]$ are numerically evaluated, respectively. The simulation result also indicates that our scheme is  more robust than the NHQC case. Notably, in the above simulations, we have set the pulse areas of all gates to be $S=\pi$, which is the best performance of NHQC but not ours. That is, $S_{min}$ in our scheme can be smaller than $\pi$, i.e., our gate can be faster than the NHQC case and the gate-fidelities present here is not the best ones for our scheme. Therefore, our scheme can implement high-fidelity, robust and low area quantum gates.

\section{Physical implementation}

In this section, we present the implementation of our proposal on   superconducting circuits and nitrogen-vacancy centres, which is directly realizable with current technology.  On superconducting circuits, we consider the transmon device, a kind of well-known artificial atom, which possesses a series of nonequal energy levels. Here, our implementation involves the three lowest levels $|g\rangle$, $|e\rangle$ and $|f\rangle$, with $|g\rangle$ and $|f\rangle$ being the qubit-states and  $|e\rangle$ is an auxiliary state \cite{Abdumalikov2013, Xu18}. Then, two microwave fields $\Omega_{l}(t)\cos(\omega_{l}t+\phi_{l})$  $(l=1,2)$ are resonantly coupled to sequential transitions of these three lowest levels, where  $\Omega_{l}(t)$, $\omega_{l}$ and $\phi_{l}$ are time-dependent driving amplitudes, time-independent driving frequencies and  phases, respectively. Setting the energy of $|g\rangle$ as the zero point, the original Hamiltonian of the system can be written as
\begin{equation}\label{Ha}
\begin{aligned}
\mathcal{H}_1 =&\omega_e|e\rangle\langle e|+\omega_f|f\rangle\langle f|\\
&+\sum_l \Omega_{l}(t)\cos(\omega_{l}t+\phi_{l})(b +b^\dag),
\end{aligned}
\end{equation}
where $b=|g\rangle\langle e|+\sqrt{2}|e\rangle\langle f|$ being the annihilation operator of the transmon. In this setting, in the interaction picture and under the rotating wave approximation, the Hamiltonian $\mathcal{H}(t)$ in Eq. (\ref{Hamiltion0}) of our proposal can be obtained. Moreover, the decay and dephasing rates of a transmon are both on the kHz level \cite{decay}, as used in our numerics.

{In addition, a Hamiltonian for holonomic two-qubit  gates is also realizable with auxiliary qubit or cavity coupled two qubits \cite{SingleloopSQ, xue1, xue2, Li}. Specifically, we consider two transmon qubits with frequencies $\omega^k_{ge}$ ($k = 0,1$) dispersively coupled to an auxiliary transmon with frequency $\omega^A_{ge}$ with the detuning being $\Delta_k =\omega^k_{ge} - \omega^A_{ge} $. Meanwhile, two qubits are respectively driven by two microwave fields $\Omega_{k}(t)\cos(\omega_{k}(t)t+\phi_{k})$ with $\omega^k_{ge}+\omega^k_{ef} - \omega^A_{ge}  \approx  \omega_k(0) $, where $\omega_{k}(t)$ is time-dependently adjusted to compensate the interaction-induced ac Stark-shifts of the transmons. In the rotating framework with respect to the driving frequency, the Hamiltonian of the $k$th qubit coupled to the auxiliary transmon can be written as
\begin{eqnarray}\label{H2q}
\mathcal{H}_0 &=&\delta_kN_k-\frac{\alpha_k}{2}(N_k-1)N_k + \delta_AN_A-\frac{\alpha_A}{2}(N_A-1)N_A,\notag\\
\mathcal{H}_I &=&g_kab^\dagger_k +\frac{\Omega_k(t)e^{i\phi_{k}}}{2}b_k+ \textrm{H.c.},
\end{eqnarray}
where $\mathcal{H}_0$ is free Hamiltonian of the coupled system, and the first two terms are  nonlinear energy levels of  $k$th transmon with its anharmonicity being $\alpha_k$, and  last two terms are  nonlinear energy levels of the auxiliary transmon with the anharmonicity being $\alpha_A$, and $\delta_k = \omega^k_{ge} - \omega_{k} $, $\delta_A = \omega^A_{ge} - \omega_{k} $, $N_k =b^\dagger_kb_k $, $N_A =a^\dagger a $, with $b_k$ and $a$ being the annihilation operator of the qubit and auxiliary transmons, respectively. $\mathcal{H}_I$  is the linear coupling term including the coupling between two transmons with strength $g_k$ and the driven induced interaction on qubit $k$ with strength $\Omega_k$. Then, in the single-excitation subspace $\mathcal{S}_1=$span$\{|fgg\rangle,|ggf\rangle,|geg\rangle\}$, where $|ijk\rangle \equiv |i\rangle\otimes|j\rangle\otimes|k\rangle$ labels product states of two qubits and auxiliary transmon, the effective Hamiltonian can be described as
\begin{equation}\label{Heff}
\begin{aligned}
\mathcal{H}_\textrm{eff} = \tilde {g}_1 e^{-i\phi_1}|fgg\rangle\langle geg| + \tilde{g}_2 e^{-i\phi_2}|ggf\rangle\langle geg| +  \textrm{H.c.},
\end{aligned}
\end{equation}
where $\tilde {g}_k = g_k\Omega_k(t)\alpha_k/[\sqrt{2}\Delta_k(\Delta_k-\alpha_k)]$. Thus, the Hamiltonian $\mathcal{H}(t)$ in Eq. (\ref{Hamiltion0}) of our proposal can be achieved in this two-qubit subspaces, and thus holonomic two-qubit  gates can also be obtained, similar to the single-qubit case.}

For a   nitrogen-vacancy centre in a diamond \cite{Zu2014}, there is a spin-triplet ground state. The Zeeman components $|m=0,\pm 1 \rangle$ induced by a magnetic field can be defined as $|m=- 1 \rangle\equiv |0\rangle$ and $|m=1 \rangle\equiv |1\rangle$ as the qubit basis  and $|m=0 \rangle\equiv |a\rangle$ as an ancillary state. The transitions from the qubit basis $|0\rangle$ and $|1\rangle$ to the ancillary state $|a\rangle$ are respectively induced by  two resonant  microwave pulses $\Omega_{j}(t)\cos(\omega_{j}t+\phi_{j})$ $ (j=1,2)$, generated by the arbitrary waveform generator. Similar to the above transmon device case, in the interaction picture and within the rotating wave approximation, the Hamiltonian $\mathcal{H}(t)$ in Eq. (\ref{Hamiltion0}) can also be  implemented.

{Meanwhile, two-qubits gates for electron and nuclear spins can also be induced with an effective three-level configuration in their coupled system \cite{Zu2014}. In this case, a nearby nuclear spin serves as the control qubit with basis vectors being $|\uparrow\rangle$ and $ |\downarrow\rangle$. By applying state-selective microwave pulses $\Omega_{i}(t)\cos(\omega_{i}t+\phi_{i})$ $ (i=1,2)$ for nuclear spin state $|\uparrow\rangle$, in the interaction picture, the coupling Hamiltonian is
\begin{equation}\label{HeffNV}
\begin{aligned}
\mathcal{H}_2 = \frac{\Omega_1(t)}{2} e^{-i\phi_1}|0\uparrow\rangle\langle a\uparrow| + \frac{\Omega_2(t)}{2} e^{-i\phi_2}|1\uparrow\rangle\langle a\uparrow| +  \textrm{H.c.},
\end{aligned}
\end{equation}
where  $|ij\rangle \equiv |i\rangle\otimes|j\rangle$ labels the product state  of the electron and nuclear spins. Thus, the Hamiltonian $\mathcal{H}(t)$ in Eq. (\ref{Hamiltion0})  can be achieved within this two-qubit subspace, and then holonomic two-qubit  gates can be implemented.}

\section{Conclusion}
In conclusion, we have proposed a fast and robust scheme to construct high-fidelity quantum gates for holonomic quantum computation based on resonant interaction of three-level quantum system via STA. In our scheme, combining the transitionless quantum driving with the LRI theory, we have inversely engineered the Hamiltonian with non-synchronous  couplings for noncyclic holonomic quantum gates. Meanwhile, by choosing a special evolution path under proper boundary conditions,  high-fidelity, robust and low area universal single-qubit gates can be constructed. Furthermore, as verified by  numerical simulations, the implemented gates have very high fidelity, and their robustness is stronger than the NHQC case, under the same conditions. Moreover, our scheme can be readily  implemented in physical systems, e.g., superconducting circuits and nitrogen-vacancy centres. Therefore, our proposal  represents a promising way towards fault-tolerant quantum computation.

\acknowledgements

This work was supported by the Key R\&D Program of Guangdong Province (Grant No. 2018B030326001), the National Natural Science Foundation of China (Grant No. 11874156),  the National Key R\&D Program of China (Grant No. 2016 YFA0301803), Guangdong Provincial Key Laboratory of Quantum Science and Engineering (Grant No. 2019B121203002), and Science and Technology Program of Guangzhou (Grant No. 2019050001).



\begin{thebibliography}{99}


\bibitem{Feynman} R. P. Feynman,
Int. J. Theor. Phys. {\bf 21}, 467 (1952).

\bibitem{berry} M. V. Berry,
Proc. R. Soc. Lond., Ser. A {\bf 392}, 45 (1984).

\bibitem{fw}
F. Wilczek and A. Zee,
Phys. Rev. Lett. \textbf{52}, 2111 (1984).

\bibitem{aa} Y. Aharonov and J. Anandan,
Phys. Rev. Lett. {\bf 58}, 1593 (1987).



\bibitem{AN1} P. Solinas, P. Zanardi, and N. Zangh\`{\i},
{Phys. Rev. A} {\bf 70}, 042316 (2004).
\bibitem{AN2} S. L. Zhu, Z. D. Wang, and P. Zanardi,
{Phys. Rev. Lett.} {\bf 94}, 100502 (2005).

\bibitem{AN3} C. Lupo, P. Aniello, M. Napolitano, and G. Florio,
{Phys. Rev. A} {\bf 76}, 012309 (2007).

\bibitem{AN4} S. Filipp, J. Klepp, Y. Hasegawa, C. Plonka-Spehr, U. Schmidt,
{Phys. Rev. Lett.} {\bf 102}, 030404 (2009).

\bibitem{AN5} J. Xu, S. Li, T. Chen, and Z.-Y. Xue,  
\textit{Front. Phys.}  {\bf15}, 41503 (2020).

\bibitem{AN6} Y. Xu, Z. Hua, T. Chen, X. Pan, X. Li, J. Han, W. Cai, Y. Ma, H. Wang, Y. Song, Z.-Y. Xue, and L. Sun, {Phys. Rev. Lett.} {\bf 124}, 230503 (2020).








\bibitem{NJP}
E. Sj\"{o}qvist, D. M. Tong, L. M. Andersson, B. Hessmo, M. Johansson, and K. Singh,
New J. Phys. \textbf{14}, 103035 (2012).

\bibitem{TongDM}
G. F. Xu, J. Zhang, D. M. Tong, E. Sj\"{o}qvist, and L. C. Kwek,
Phys. Rev. Lett. \textbf{109}, 170501 (2012).


\bibitem{surface1} Y.-C. Zheng and T. A. Brun,
 Phys. Rev. A {\bf 91}, 022302 (2015).

\bibitem{Singleloopxu} G. F. Xu, C. L. Liu, P. Z. Zhao, and D. M. Tong,
Phys. Rev. A {\bf 92}, 052302 (2015).

\bibitem{Singleloop}
E.~Herterich and E.~Sj\"{o}qvist,
Phys. Rev. A \textbf{94}, 052310 (2016).


\bibitem{SingleloopSQ}
Z.-P. Hong, B.-J. Liu, J.-Q. Cai, X.-D. Zhang, Y. Hu, Z. D. Wang, and Z.-Y. Xue,
Phys. Rev. A \textbf{97}, 022332 (2018).


\bibitem{surface2} J. Zhang, S. J. Devitt, J. Q. You, and F. Nori,
Phys. Rev. A \textbf{97}, 022335 (2018).


\bibitem{ChenAn} T. Chen, J. Zhang, and Z. Y. Xue,
Phys. Rev. A \textbf{98}, 052314 (2018).


\bibitem{eric} N. Ramberg and E. Sj\"{o}qvist,
  Phys. Rev. Lett. {\bf 122}, 140501 (2019).

\bibitem{surface3} C. Wu, Y. Wang, X.-L. Feng, and J.-L. Chen,
Phys. Rev. Appl. {\bf 13}, 014055 (2020).



\bibitem{Abdumalikov2013} A.~A. Abdumalikov, J.~M. Fink, K.~Juliusson, M.~Pechal, S.~Berger, A.~Wallraff, and S.~Filipp,  
    Nature (London) \textbf{496}, 482 (2013). 

\bibitem{Feng2013} G.~Feng, G.~Xu, and G.~Long,
Phys. Rev. Lett. \textbf{110}, 190501   (2013).


\bibitem{Zu2014} C.~Zu, W.-B. Wang, L.~He, W.-G. Zhang, C.-Y. Dai, F.~Wang, and L.-M. Duan,
Nature (London) \textbf{514}, 72 (2014).   

\bibitem{Arroyo-Camejo2014} S.~Arroyo-Camejo, A.~Lazariev, S.~W. Hell, and G.~Balasubramanian,
Nat. Commun. \textbf{5}, 4870 (2014).


\bibitem{nv2017} Y. Sekiguchi, N. Niikura, R. Kuroiwa, H. Kano, and H. Kosaka,
Nat. Photonics {\bf 11}, 309 (2017). 

\bibitem{nv20172}
B. B. Zhou, P. C. Jerger, V. O. Shkolnikov, F. J. Heremans, G. Burkard, and D. D. Awschalom,
Phys. Rev. Lett.  {\bf 119}, 140503 (2017).


\bibitem{li2017}  H. Li, L. Yang, and G. Long,
Sci. China: Phys., Mech. Astron. {\bf 60}, 080311(2017).

\bibitem{Xu18}
Y. Xu, W. Cai, Y. Ma, X. Mu, L. Hu, T. Chen, H. Wang, Y.-P. Song, Z.-Y. Xue, Z.-Q. Yin, and L. Sun,
Phys. Rev. Lett. {\bf 121}, 110501 (2018).

\bibitem{ni2018}
N. Ishida, T. Nakamura, T. Tanaka, S. Mishima, H. Kano, R. Kuroiwa, Y. Sekiguchi, and H. Kosaka,
Opt. Lett. {\bf 43}, 2380 (2018).

\bibitem{kn2018} K. Nagata, K. Kuramitani, Y. Sekiguchi, and H. Kosaka,
Nat. Commun. {\bf 9}, 3227 (2018).

\bibitem{dje2019} D. J. Egger, M. Ganzhorn, G. Salis, A. Fuhrer, P. Muller, P. K. Barkoutsos, N. Moll, I. Tavernelli, and S. Filipp,
Phys. Rev. Appl. {\bf 11}, 014017 (2019).





\bibitem{Zheng16} S. B. Zheng, C. P. Yang, and F. Nori,
Phys. Rev. A {\bf 93}, 032313 (2016).

\bibitem{Jing17}  J. Jing, C.-H. Lam, and L.-A. Wu,
Phys. Rev. A {\bf 95}, 012334 (2017).




\bibitem{encode}  Y.-C. Zheng and T. A. Brun,
Phys. Rev. A {\bf 89}, 032317 (2014).

\bibitem{zhangj2014}
J. Zhang, L.-C. Kwek, E. Sj\"{o}qvist, D. M. Tong, and P. Zanardi,
Phys. Rev. A {\bf 89}, 042302 (2014).

\bibitem{liang2014}  Z.-T. Liang, Y.-X. Du, W. Huang, Z.-Y. Xue, and H. Yan,
Phys. Rev. A {\bf 89}, 062312 (2014).


\bibitem{zhouj} J. Zhou, W.-C. Yu, Y.-M. Gao, and Z.-Y. Xue,
Opt. Express {\bf 23}, 14027 (2015).

\bibitem{xue1} Z.-Y. Xue, J. Zhou, and Z. D. Wang,
Phys. Rev. A {\bf 92}, 022320 (2015).

\bibitem{wangym2016} Y. Wang, J. Zhang, C. Wu, J. Q. You, and G. Romero,
Phys. Rev. A {\bf 94}, 012328 (2016).

\bibitem{xue2} Z.-Y. Xue, J. Zhou, Y.-M. Chu, and Y. Hu,
Phys. Rev. A {\bf 94}, 022331 (2016).


\bibitem{xue3} Z.-Y. Xue, F.-L. Gu, Z.-P. Hong, Z.-H. Yang, D.-W. Zhang, Y. Hu, and J. Q. You,
Phys. Rev. Appl. {\bf 7}, 054022 (2017).

\bibitem{zhaopz2017} P. Z. Zhao, G. F. Xu, Q. M. Ding, E. Sj\"{o}qvist, and D. M. Tong,
  Phys. Rev. A {\bf 95}, 062310 (2017).

\bibitem{xue4} L.-N. Ji, T. Chen, and Z.-Y. Xue,
Phys. Rev. A {\bf 100}, 062312 (2019).


\bibitem{wangym2020} Y. Wang, Y. Su, X. Chen, and C. Wu,
Phys. Rev. Appl. {\bf 14}, 044043 (2020).


\bibitem{composite} G. F. Xu, P. Z. Zhao, T. H. Xing, E. Sj\"{o}qvist, and D. M. Tong,
Phys. Rev. A \textbf{95}, 032311 (2017).

\bibitem{zhu2019}
Z. Zhu, T. Chen, X. Yang, J. Bian, Z.-Y. Xue, and X. Peng,
Phys. Rev. Appl. \textbf{12}, 024024 (2019).

\bibitem{dd} Y. Sekiguchi, Y. Komura, and H. Kosaka,
 Phys. Rev. Appl. {\bf 12}, 051001 (2019).

\bibitem{liubj17}
B.-J. Liu, Z.-H. Huang, Z.-Y. Xue, and X.-D. Zhang,
Phys. Rev. A \textbf{95}, 062308 (2017).

\bibitem{Liu18}
B.-J. Liu, X.-K. Song, Z.-Y. Xue, X. Wang, and M.-H.Yung,
Phys. Rev. Lett. \textbf{123}, 100501 (2019).

\bibitem{yan2019}
T. Yan, B.-J. Liu, K. Xu, C. Song, S. Liu, Z. Zhang, H. Deng, Z. Yan, H. Rong, K. Huang, M.-H. Yung, Y. Chen, and D. Yu,
Phys. Rev. Lett. \textbf{122}, 080501 (2019).

\bibitem{Li}
S. Li, T. Chen, and Z.-Y. Xue,
Adv. Quantum Technol. \textbf{3}, 2000001 (2020).

\bibitem{ai2020} M.-Z. Ai, S. Li, Z. Hou, Z.-Y. Xue, J.-M. Cui, Y.-F. Huang, C.-F. Li, and G.-C. Guo,
 Phys. Rev. Appl. {\bf 14}, 054062 (2020).

\bibitem{ai2021} M.-Z. Ai, S. Li,  Z.-Y. Xue, J.-M. Cui, Y.-F. Huang, C.-F. Li, and G.-C. Guo,
arXiv:2101.07483.

%

\bibitem{GPC}  B.-J. Liu, Y.-S. Wang, and M.-H. Yung,
arXiv:2008.02176.

\bibitem{DCGLi} S. Li and Z.-Y. Xue,
arXiv:2012.09034.


\bibitem{xugf2018} G. F. Xu, D. M. Tong, and E. Sj\"{o}qvist,
Phys. Rev. A 98, 052315 (2018).

\bibitem{zhang2019} F. Zhang, J. Zhang, P. Gao, and G. Long,
 Phys. Rev. A {\bf 100}, 012329 (2019).


\bibitem{Chentoc3} T. Chen, P. Shen, and Z.-Y. Xue,
Phys. Rev. Appl. {\bf 14}, 034038 (2020).

\bibitem{BNHQC}
B.-J. Liu, Z.-Y. Xue, and M.-H.Yung,
arXiv:2001.05182.

\bibitem{yuyang} Z. Han, Y. Dong, B. Liu, X. Yang, S. Song, L. Qiu, D. Li, J. Chu, W. Zheng, J. Xu, T. Huang, Z. Wang, X. Yu, X. Tan, D. Lan, M.-H. Yung, and  Y. Yu,
arXiv:2004.10364.


\bibitem{STA2010b} X. Chen, A. Ruschhaupt, S. Schmidt, A. del Campo, D. Gu\'{e}ry-Odelin, and  J. G. Muga,
 Phys. Rev. Lett. {\bf 104}, 063002 (2010).

\bibitem{STA2019} D. Gu\'{e}ry-Odelin, A. Ruschhaupt, A. Kiely, E. Torrontegui, S. Mart\'{\i}anez-Garaot, and J. G. Muga,
Rev. Mod. Phys. {\bf 91}, 045001 (2019).

\bibitem{LR} H. R. Lewis, and W. B. Riesenfeld,
J. Math. Phys. {\bf 10}, 1458 (1969).

\bibitem{Chen12}
X. Chen and J. G. Muga,
 Phys. Rev. A {\bf 86}, 033405 (2012).


\bibitem{xia2018} 
Y.-H. Chen, Z.-C. Shi, J. Song, and Y. Xia
Phys. Rev. A {\bf 97}, 023841 (2018).


\bibitem{zhou19}J. Zhou, S. Li, T. Chen, and Z.-Y. Xue,
Ann. Phys. (Berlin) {\bf 531}, 1800402 (2019).

\bibitem{Yan19} Y. Yan, Y. Li, A. Kinos, A. Walther, C. Y. Shi, L. Rippe, J. Moser, S. Kr\"{o}ll, and X. Chen,
Opt. Express {\bf 27}, 008267 (2019).




\bibitem{IR2020} T. Chen and Z.-Y. Xue,
Phys. Rev. Appl. {\bf 14}, 064009 (2020).

\bibitem{Cost1} Y. J. Zheng, S. Campbell, G. D. Chiara, and D. Poletti,
Phys. Rev. A {\bf 94}, 042132 (2016).

\bibitem{Cost2}O. Abah, R. Puebla, A. Kiely, G. D. Chiara, M. Paternostro, and S. Campbell, 
New J. Phys. {\bf 21}, 103048 (2019).

\bibitem{Poyatos97} J. F. Poyatos, J. I. Cirac, and P. Zoller,
Phys. Rev. Lett. {\bf78}, 390 (1997).


\bibitem{decay} M. J. Peterer, S. J. Bader, X. Jin, F. Yan, A. Kamal, T. J.
Gudmundsen, P. J. Leek, T. P. Orlando, W. D. Oliver, and S.
Gustavsson, Phys. Rev. Lett. {\bf 114}, 010501 (2015).

\end{thebibliography}
\end{document}